\documentclass{mem}
\usepackage{natbib}\usepackage{txfonts}\usepackage{balance}
\usepackage{graphicx}
\usepackage[a4paper]{hyperref}
\idline{75}{282}
\begin{document}
\def\teff{$T\rm_{eff }$}
\def\kms{$\mathrm {km s}^{-1}$}

\title{Evolution and seismology of $\alpha$ Centauri}

   \subtitle{}

\author{
J. \,Montalb\'an
\and A. \,Miglio
          }

  \offprints{J. Montalb\'an}

\institute{Institut d'Astrophysique et de G\'eophysique de l'Universit\'e de Li\`ege,
All\'ee du 6 Ao\^ut, 17 B-4000 Li\`ege, Belgium
\email{j.montalban@ulg.ac.be}
}

\authorrunning{Montalb\'an \& Miglio }


\abstract{
Solar-like oscillations detected in both components of the binary system $\alpha$ Centauri provide strong constraints on the fundamental parameters of the stellar system. We model $\alpha$ Centauri by means of a Levenberg-Marquardt minimization algorithm including seismic and classical constraints. 
Computations, that were perfomed decreasing significanly the weight of $\alpha$Cen~B seismic data in the calibration procedure, 
predict  small separations in good agreement with new observations of solar-like oscillations in $\alpha$Cen~B  by Bedding  (these proceedings).
}
\maketitle{}

\section{Modelling $\alpha$ Centauri AB}
The numerous and precise observational constraints, including solar-like oscillations, make the binary system $\alpha$ Centauri a suitable target to test stellar structure and evolution models in conditions that are slightly different than in the Sun. This is the reason why $\alpha$ Cen has been the subject of many theoretical studies, in particular since  \citet{Bouchy02}\  detected and identified  solar-like oscillations 
in $\alpha$ Cen A (see e.g. \citet{Thevenin02}, \citet{Thoul03} and \citet{Eggenberger04}).

We model $\alpha$ Cen using a non-linear fitting algorithm that performs a simultaneous least-square adjustment of all the observables, both classical and  seismic. Our aim 
is to study how the fitted parameters (the ``best model'') depend on the choice of classical and seismic observables included in the fitting procedure, and on the ``physics'' included in stellar models.
The asteroseismic data for component A and B are taken respectively from \citet{Bouchy02} and \citet{Carrier03} and are included in the calibration by means of the large separation $\Delta\nu$, the small separation $\delta\nu$ and the combination of frequencies $r_{\rm 02}$ introduced by \citet{Roxburgh03}. The whole set of models is computed using CLES (Code Li\'egeois d'Evolution  Stellaire). A detailed description of the fitting method and of the results of our study is presented in \citet{Miglio05}.
\begin{figure*}[t!]
\resizebox{0.5\hsize}{!}{\includegraphics[clip=true]{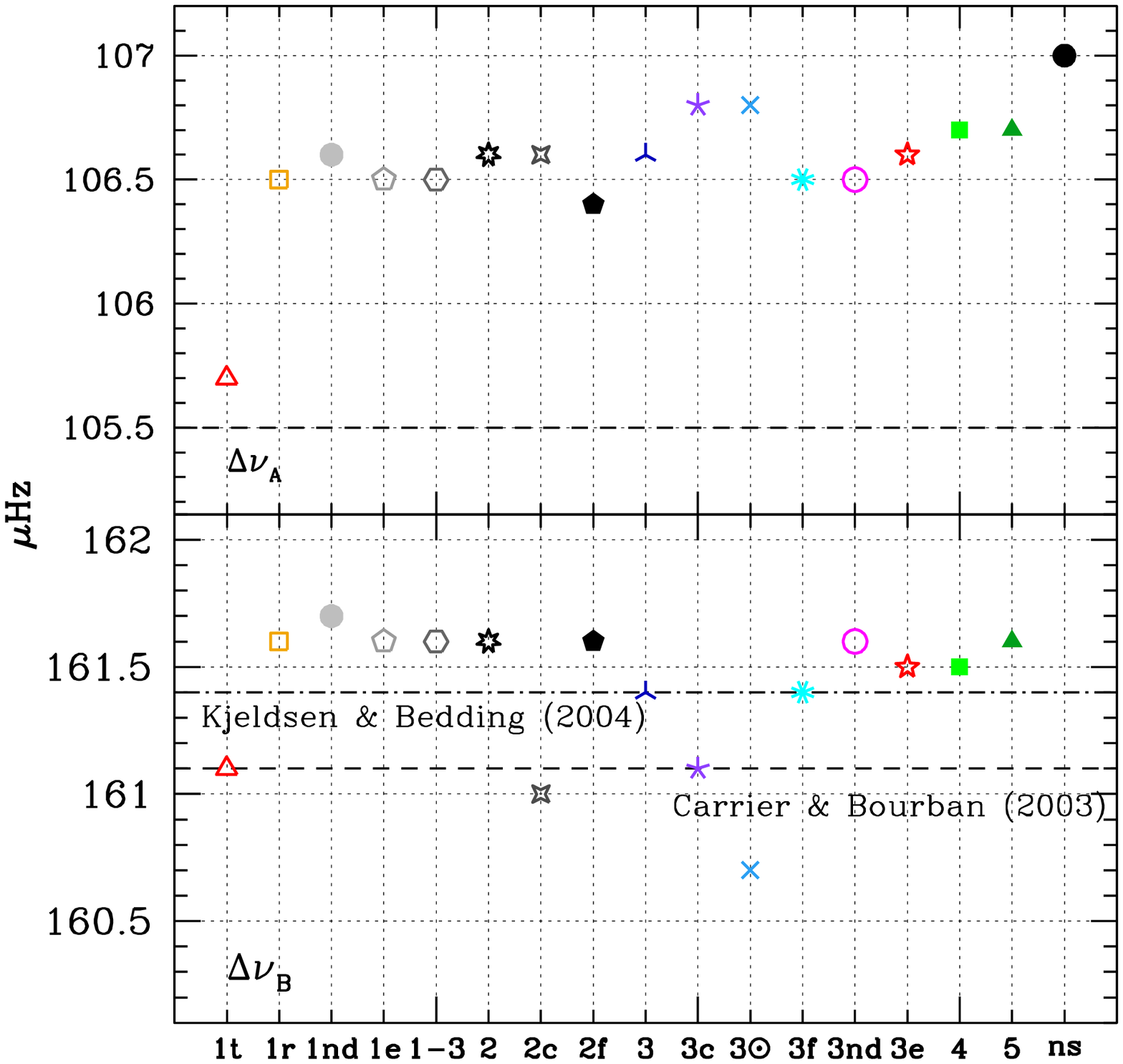}}
\resizebox{0.5\hsize}{!}{\includegraphics[clip=true]{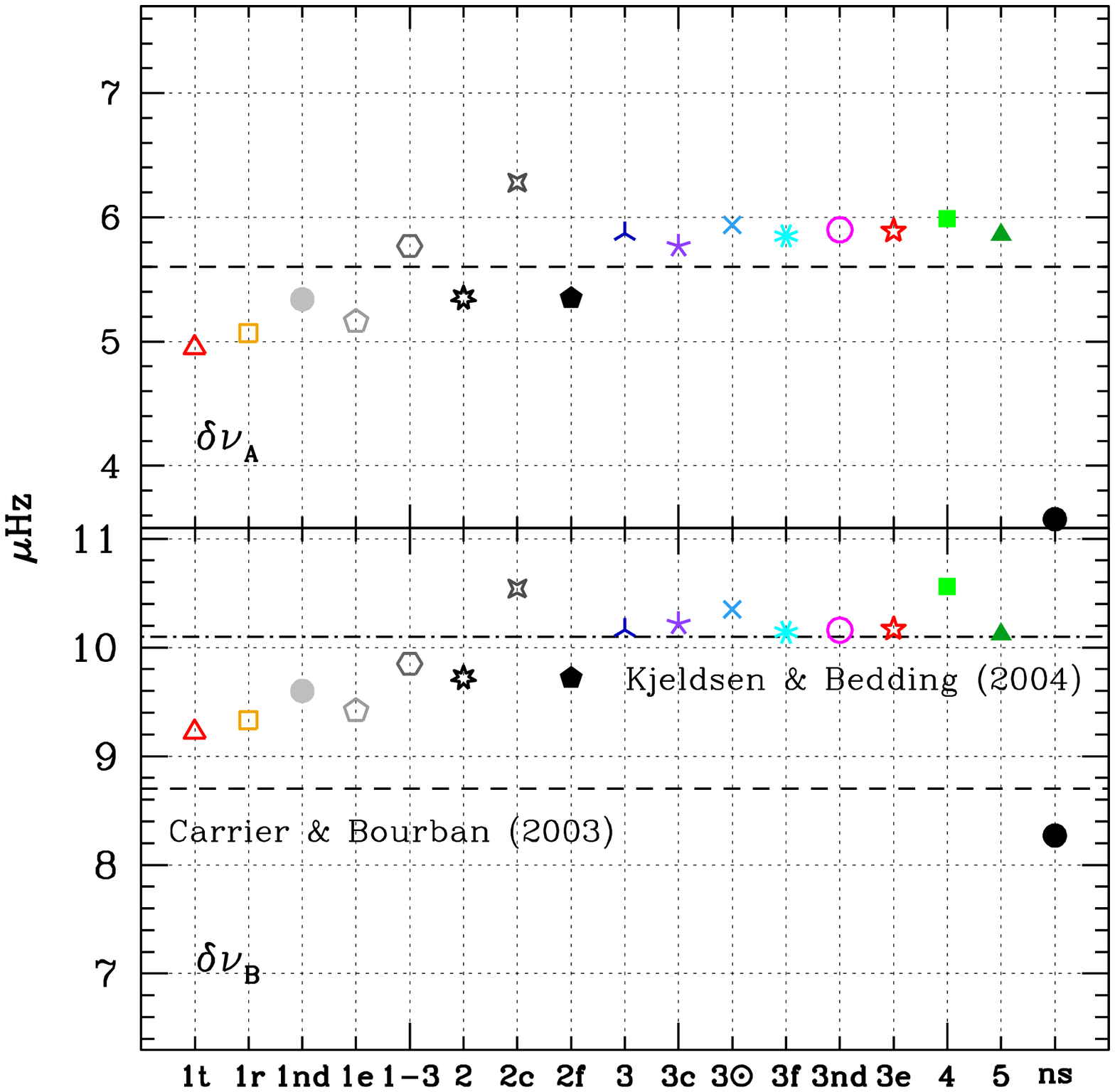}}
\caption{\footnotesize
 Averaged large and small frequency separations of the fitted models. Meaning of numeric labels: 
1 (fixed masses and $\Delta\nu$, $\delta\nu$ as seismic constraints); 
2 (as 1 but with variable masses); 3 (as 2 but using $r_{02}(n)$ of component A as seismic constraint);
4 (as 3 but including convective overshooting) and 
5 (as 3 but using Asplund et al. (2005) instead of Grevese \& Noels (1993)).
1-3 (as 3 but with fixed mass)
Meaning of  alphabetic labels: t (effective temperature \citep{Neuforge-Verheecke97} instead of radii \citep{Kervella03} included in the constraints); nd (models without diffusion); f (FST convection treatment); 
e (CEFF equation of state instead of OPAL2001);
c (a unique mixing-length parameter) and 
ns (fit without seismic constraints)
}
\label{fig:separations}
\end{figure*}

We find that the age of the system determined by the calibration is deeply biased by the small frequency separation of component B. In some of our calibrations (see Fig. \ref{fig:separations}), where we decide not to include $\delta\nu_{\rm B}$ in the fitting, we find a lower value of the age ($\sim$ 5.8 instead of $\sim$ 6.4 Gyr). In these calibrations the value of $\delta\nu_{\rm B}$ predicted by our models is significantly higher than the one given in \citet{Carrier03} but in very good agreement with the value determined by new observations (see Bedding, these proceedings), as shown in the lower-right panel of Fig. \ref{fig:separations}.

Solar-like oscillations provide stringent constraints to the modelling of the binary system $\alpha$ Centauri. Nonetheless,
the precision of the current seismic data can lead to some bias in the determination of fundamental parameters of the system,
and does not allow to discriminate between different physics.
Solar-like oscillation frequencies determined by long and uninterrupted observations would also give us the opportunity to infer properties of the stellar structure that are otherwise inaccessible.

\begin{acknowledgements}
J.M and A.M acknowledge financial support from the Prodex-ESA Contract 15448/01/NL/Sfe(IC). A.M. is also thankful to the meeting organizers
 for the financial support.
\end{acknowledgements}

\bibliographystyle{aa}

\end{document}